# A competitive binding model predicts nonlinear responses of olfactory receptors to complex mixtures

Vijay Singh[a,b], Nicolle R. Murphy[c], Vijay Balasubramanian[a,b,1,2], and Joel D. Mainland[c,d,2]

[a]Computational Neuroscience Initiative, University of Pennsylvania, Philadelphia, PA, 19104, USA; [b]Department of Physics, University of Pennsylvania, Philadelphia, PA, 19104, USA; [c]Monell Chemical Senses Center, Philadelphia, PA, 19104, USA; [d]Department of Neuroscience, University of Pennsylvania, Philadelphia, PA, 19104, USA

**In color vision, the quantitative rules for mixing lights to make a target color are well understood. By contrast, the rules for mixing odorants to make a target odor remain elusive. A solution to this problem in vision relied on characterizing receptor responses to different wavelengths of light and subsequently relating these responses to perception. In olfaction, experimentally measuring receptor responses to a representative set of complex mixtures is intractable due to the vast number of possibilities. To meet this challenge, we develop a biophysical model that predicts mammalian receptor responses to complex mixtures using responses to single odorants. The dominant nonlinearity in our model is competitive binding (CB): only one odorant molecule can attach to a receptor binding site at a time. This simple framework predicts receptor responses to mixtures of up to twelve monomolecular odorants to within 15% of experimental observations and provides a powerful method for leveraging limited experimental data. Simple extensions of our model describe phenomena such as synergy, overshadowing, and inhibition. We demonstrate that the presence of such interactions can be identified via systematic deviations from the competitive binding model.**

Olfaction | Receptors | Odor | Odorant mixtures | Competitive binding

In the field of flavors and fragrances, methods for mixing odorants to make a target odor are largely the domain of experts who have undergone years of training. Their expertise comes from examining historical formulae as well as extensive trial-and-error work, and their methods are primarily qualitative. In vision, by contrast, the rules for mixing lights to make a target color are quantitative and well-developed. These rules are derived from a detailed characterization of human color perception and its relation to cone photoreceptor spectral sensitivities (1–3). Indeed, known tuning curves relate the wavelength of light to the responses of three types of cone photoreceptors. These input-response functions are then incorporated into models that extrapolate from the responses to single wavelengths to an arbitrary mixture of wavelengths. Finally, these receptor responses are used to predict color perception.

Here, we propose an analogous approach for characterizing the response of receptors to single odorants and modeling the responses to combinations of odorants. Simple summation models are widely used (4–8), but fail to account for several observed interactions, such as suppression, masking, hyperadditivity (or synergy), hypoadditivity (or compression), configural perception, and overshadowing. The wide variety of mixture interactions suggests that a simple model would struggle to explain experimental results, but here we show that a minimal biophysical description of odorant-receptor interaction incorporating the simplest possible nonlinearity, namely competition between molecules for the binding site, can successfully predict the responses of many mammalian odor receptors to complex molecular mixtures. Previously, Rospars et al. (2008) (9) found that responses of olfactory receptor neurons to some simple binary mixtures were largely consistent with a similar model, and could display both hyper- and hypo-additivity. Related results for binary mixtures have also been reported for neurons in the accessory olfactory system (10), and in the antennal lobes of Drosophila (11) and locust (12). Cruz and Lowe (2013) (13) subsequently developed a biophysically motivated version of this model and applied it to glomerular imaging. Marasco et al. (2016) (14) extended this work to allow different odorants to have different Hill coefficients, and thus different degrees of binding cooperativity, which allowed for the phenomena of synergy and inhibition, although a biophysical motivation was lacking.

Here, we present two key steps forward. First, we collect receptor data for a large set of odors and show that our competitive binding model largely accounts for the response of olfactory receptors to complex mixtures of up to 12 odorants. Second, we develop a systematic strategy to identify additional nonlinear interactions among odorants and receptors that go beyond the effects of competitive binding. Our approach is rooted in basic biophysics. For example, the extended models consider consequences of known phenomena like receptors with multiple binding sites, facilitation by already bound odorants, non-competitive inhibition, and heterodimerization of odorant molecules in mixture, and predict effects such as synergy, antagonism (15) and overshadowing (16) in receptor responses. Such phenomena are reported in studies of human olfactory perception (17), but their origin is unknown. We hypothesize that such nonlinear effects, previously assumed to be of neural origin, may already have a contribution from interactions at the level of the receptor.

## Results

**Competitive binding model.** The response of a receptor to an odor can be modeled in terms of the binding and unbinding of odorant molecules to the receptor binding site. We assume that only one molecule can attach to a binding site at a time, leading to competition. In the presence of many odorants, the outcome of competition depends on three parameters: the concentration of the individual molecules, the efficacy with which the molecule activates the receptor, and the affinity of the molecule for the binding site.

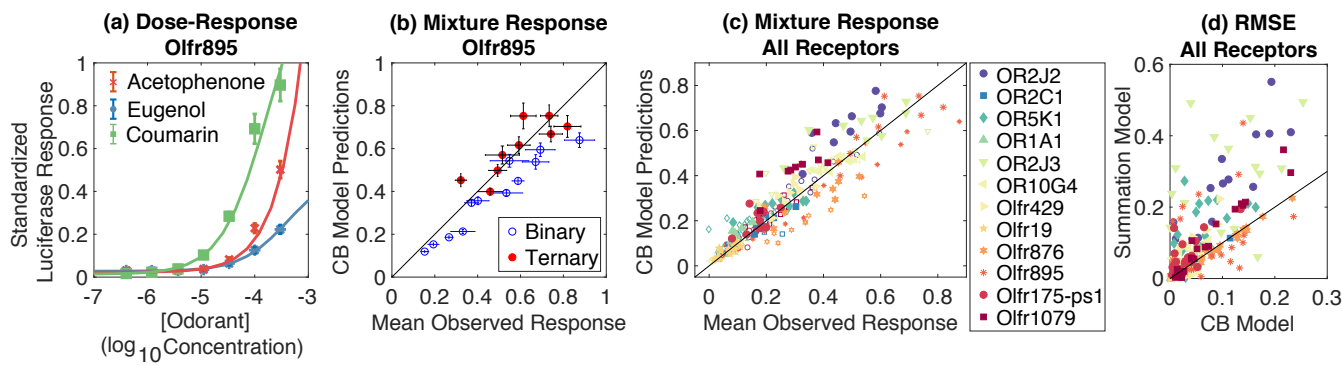

**Fig. 1.** A competitive binding model predicts olfactory receptor response to binary/ternary mixtures: (a) Response of receptor Olfr895 to individual odorants. Markers show mean experimental response ± one standard deviation (s.d.). Smooth curves show the competitive binding (CB) model with parameters chosen to minimize the error $E_i$ defined as the root mean squared error between model and data weighted by experimental standard deviations (see text and SI Appendix, Model parameter estimation). (b) Response of Olfr895 to binary and ternary mixtures. CB model predictions plotted against experimental responses averaged over 4 replicates. The diagonal line in black is the unit slope line. Horizontal bars represent ± one s.d. Vertical error bars are s.d. over mixture predictions for 300 randomly chosen sets of model parameters constrained so that the error $E_i$ was lower than $\lceil E_i^{min} \rceil$, where $E_i^{min}$ is the error for the best fit parameters and $\lceil\ \rceil$ is the ceiling function (see SI Appendix, Variance in CB model predictions). In general this amounts to picking random parameter sets such that the model dose-response curves lie within a standard deviation of the experimental mean (see SI Appendix: Methods). (c) Response of 12 olfactory receptors from humans and mice to binary and ternary mixtures (CB model vs. average experimental responses; binary mixture responses = open markers; ternary mixture responses = filled markers; diagonal line = unit slope line). For these 12 receptors, the median root mean square error (RMSE) was below 0.1. (See SI Appendix, Fig. S1 for alternative measures of prediction error.) (d) RMSE of summation model plotted vs the RMSE of CB model. RMSE of summation model lies above the diagonal unit slope line for most mixtures, indicating that the summation model predictions are worse compared to the CB model.

We modeled the response of a receptor to the binding of an odorant as a two-step process (see SI Appendix, Competitive binding model, (18)). Such models have widely been used to study kinetics of chemical and biological systems starting with Michelis and Menten in 1913 (19–23). In the first step, the molecule binds reversibly to the binding site. At this stage, the bound receptor can either dissociate, giving back the odorant and the unbound receptor, or can reversibly go to an active state. The transition to the active state is the second step. In the active state, the odorant-receptor complex elicits a detectable response. In our experiments, this response is measured using a luciferase reporter in a cell-based assay (24).

In this competitive binding (CB) model, the response of a receptor $F(\{c_i\})$ to a mixture of N odorants with concentrations represented by $\{c_i\}$ is given by (derivation in SI Appendix, Competitive binding model):

$$F(\{c_i\}) = \frac{F_{\max} \sum_{i=1}^{N} \frac{e_i c_i}{EC50_i}}{\left(1 + \sum_{i=1}^{N} \frac{c_i}{EC50_i}\right)} \quad [1]$$

Here, $EC50_i$ is the concentration at which the response is half of the maximum for odorant $i$, $e_i$ is the efficacy of the receptor for odorant $i$, and $F_{\max}$ parameterizes the total receptor concentration and overall response efficiency (see SI Appendix).

**CB model predicts receptor responses to mixtures.** We used a heterologous assay to measure receptor responses to three monomolecular odorants (eugenol, coumarin, and acetophenone) known to broadly activate mammalian odor receptors (25). Dose-response curves were measured for 15 receptors (e.g., Fig. 1a) by stimulating the receptors across the full range of concentrations allowed by our assay ([0,0.3 mM]; Methods). These 15 receptors were then stimulated with 21 mixtures (12 binary, 9 ternary) of eugenol, coumarin and acetophenone (Methods; SI Appendix, Table 1) with concentrations now chosen to avoid receptor saturation.

We first fit the CB model to the dose-response data for individual odorants (N=1 in Eq. 1). We selected parameters to minimize the root mean squared error between predictions and measurements (SI Appendix, Table 4) weighted by the experimental standard deviation (Methods; example in Fig. 1a, further details in SI Appendix, Model parameter estimation). The parameters that best reproduced the single-odorant data were then used to predict the response to odorant mixtures (Fig. 1b,c).

For most receptors (12 out of 15), the root mean squared error (RMSE; see Methods and SI Appendix, Fig. S1) was low (median below 0.1), and small relative to the observed response (median of RMSE/Observed Response = 0.16) and compared to the experimental standard deviations (median RMSE/STD = 1.2). (See next section for the remaining 3 receptors.) The results are consistent with the hypothesis that the receptor response is generated by the CB model (chi-squared test, null hypothesis that CB model generates the responses is not rejected, p>0.999; details in SI Appendix). We also tested that the CB model predictions are robust to parameter variations that keep the predicted dose-response curves within a standard deviation of the best fit (SI Appendix, Fig. S1).

Next we compared the RMSE of the CB model to that of a summation model where responses were predicted to be linear sums of responses to individual odorants in the dose-response analysis (See Material and Methods). Such summation models have previously been applied to the responses of Olfactory Sensory Neurons and in the olfactory bulb (4–7). In addition, the human psychophysics literature frequently assumes a summation model as the default for the perceived intensity of binary mixtures (8, 26). We found that the RMSE for the CB model was lower than the summation model (Fig. 1(d)). This improvement occurred even though mixture concentra-tions were chosen to lie in an approximately "linear" regime that avoided saturation. We also tested that the CB model predictions were better than the summation model in terms of other measures of prediction error (SI Appendix, Fig. S2a-b). The median CB model predictions lie within ∼ 15% of the actual magnitude of the response to individual mixtures. These results confirm the model's accuracy.

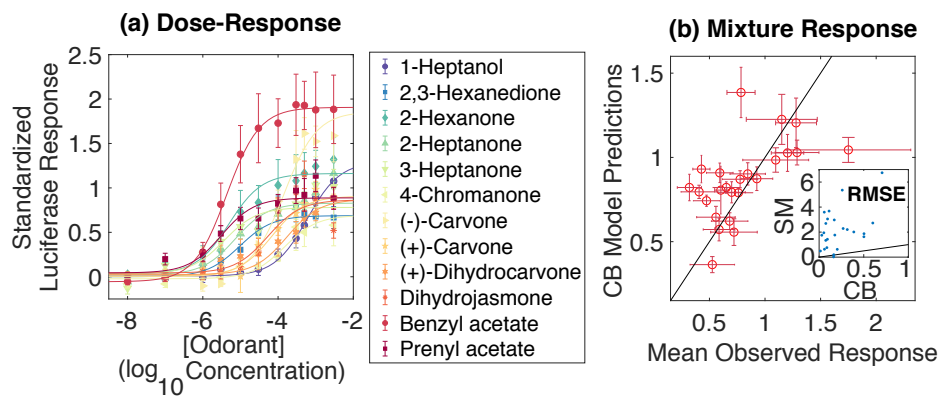

**Fig. 2. A competitive binding model predicts olfactory receptor response to 12-component mixtures** (a) Response of receptor Olfr168 (mouse) to 12 individual odorants. Markers show mean experimental measurements ± one s.d. Smooth curves show CB model. (b) CB model predictions v/s experimental responses for Olfr168. The errorbars represent ± one s.d. (Inset) RMSE of summation model (SM) plotted vs the CB model. The diagonal line in black is the unit slope line.

To further challenge the model, we studied the response of olfactory receptors to mixtures that were more comparable in complexity to natural odors, which typically have about 3-40 perceptually important components (27, 28). We focused on mouse receptor Olfr168, which responds to a large number of odorants (25). From the data in Saito et al. (2009) we identified 12 odorants that evoked responses in this receptor (Methods). Similar to the procedure above, we first fit the dose-response measurements for all 12 odorants to get the best parameters for the receptor (Fig. 2a, SI Appendix, Table 5). Then, we used the competitive binding model to predict receptor responses to mixtures with all 12 odorants present in diverse proportions (Methods; Fig. 2b). Trivially, a combination of many odors at a moderate concentration will activate receptors to saturation. To avoid this, we chose concentrations of the mixture components such that the receptor activation in response to the full mixture was above threshold and below saturation. The model predicted the receptor responses to such complex mixtures very well (Olfr168: CB median RMSE = 0.16). The results are consistent with the hypothesis that the receptor response is generated by the CB model (chi-squared test, null hypothesis that CB model generates the responses is not rejected, p>0.999; details in SI Appendix). The CB model also outperformed a summation model of mixture response by more then 10-fold (Olfr168: Summation median RMSE = 1.91). Thus, for complex odor mixtures such as those occurring naturally, our nonlinear competitive binding model presents a dramatic improvement over a summation model.

We wondered whether the specificity of receptor-odorant interactions determines model accuracy, or whether good prediction results from simply fixing responses to be sigmoidal in the response range of a typical receptor. To test this, we compared the CB model to a shuffled model where, instead of using the specific dose-response curves of mixture components, we selected dose-response parameters randomly from all such parameters available in our dataset (Methods), and averaged the prediction error over 300 such random choices. The competitive binding model outperformed the shuffled model for both binary-ternary mixtures (Shuffled RMSE = 10 to 100 times CB RMSE; SI Appendix, Fig. S3c) and the 12-component mixtures (Shuffled median RMSE = 0.95 ∼ 6 times CB median RMSE).

**Extensions of the model.** So far, we have considered the simplest possible form of odorant-receptor interaction: only one odorant molecule binds a receptor binding site at a time. Surprisingly, most of the receptors studied in our experiments were well-described by this model. Competitive binding can produce essentially three types of nonlinear receptor responses to presentation of mixtures (Fig. 3a-c): (1) Domination by the odorant that gives the highest response individually (overshadowing, Fig. 3a), (2) A response in between those to the individual odorants (suppression, Fig. 3b), (3) Domination by the odorant that gives the lowest individual response (also called overshadowing, Fig. 3c). These effects can arise both from the intrinsic properties of the receptor-odorant interaction (difference in EC50) or due to extrinsic factors such as the ratio of concentrations. Such qualitative effects have been reported previously (14) in a phenomenological model that has a more complex form of response to mixtures. We have shown here that these effects can already be exhibited by a simple model directly rooted in biophysical competition between the odorant molecules seeking to occupy the receptor.

Our model can be easily extended to incorporate additional biophysical interactions that produce effects such as synergy (17) and inhibition (29). Although previous work (9, 13, 14) has explored possible mathematical functions that can be used to fit such nonlinearities in receptor response data, a biophysical understanding of the origin of these effects has been missing. Some recent progress on this front is reported by (30) who focused on antagonism in receptors, and proposed, e.g., additional interactions with cell membranes as a mechanism for non-specific suppression. These authors also argued on theoretical grounds that antagonism can normalize receptor neuron population activities, improving the performance of decoders of the response ensemble. Our approach of starting from the simplest interactions at the molecular level provides an avenue for systematically identifying important interactions. For example, consider facilitation, where the binding of an odorant promotes the binding of other odorants to the same site. Such an interaction modifies Eq. 1 (see Methods and SI Appendix, Facilitation) and produces effects such as synergy (Fig. 3d), in which the response of the receptor is higher than the sum of the response to both individual odors, and inhibition (Fig. 3e) where the response is below the response to both individual odorants. This is in addition to the effects already produced by competitive binding (overshadowing and suppression, Fig. 3a-c). Alternatively, if there are multiple independent binding sites for odorants, the mixture response will be the sum of the individual components (SI Appendix, Independent binding sites). More complex biophysical interactions, such as non-competitive inhibition (SI Appendix, Non-competitive inhibition), hetero-dimerization (SI Appendix, Odorant dimerization), catalysis by odor molecules, etc. can similarly be added to the basic model in a principled way.

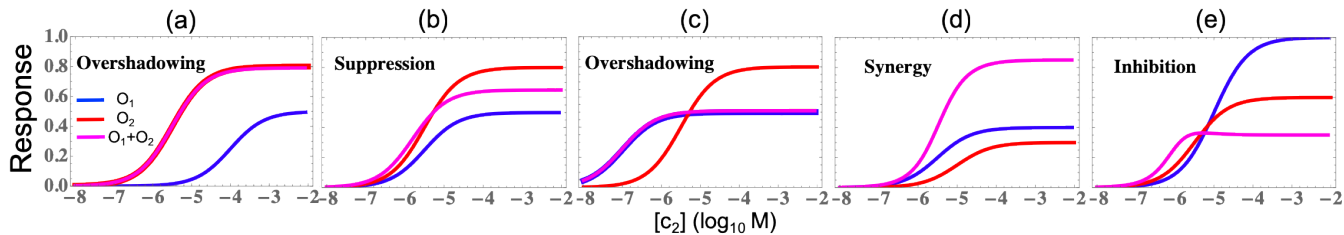

**Fig. 3. Phenomena exhibited by the competitive binding model:** The competitive binding model with different parameter choices shows diverse effects for binary mixtures (purple) of two odorants (red and blue). Shown here are effects due to variations of $EC50$: (a) overshadowing by the odorant with the higher individual response ($EC50_1 = 10^{-4.0}$), (b) suppression, where the response is in between the responses of the individual odorants ($EC50_1 = 10^{-5.5}$), and (c) overshadowing by the receptor that produces the lower individual response ($EC50_1 = 10^{-7.0}$). Value of the other model parameters: $F_{max} = 1$, $e_1 = 0.5$, $e_2 = 0.8$ and $EC50_2 = 10^{-5.5}$. We assume equimolar mixtures ($c_1/c_2 = 1$). **Phenomena exhibited by the extended model including odorant facilitation.** Facilitation of odorant binding by another odorant molecule in mixture leads to additional effects like synergy and inhibition. (d) Synergy: receptor response is higher than response to both the individual odorants. (e) Inhibition: response to mixtures is lower than the response to either individual odorant. Functional form for facilitation and parameter choices leading to synergy and inhibition are given in SI Appendix (SI Appendix: Facilitation).

To illustrate our proposed systematic approach to adding interactions, we considered the three receptors whose responses to binary and ternary mixtures deviated significantly from the predictions of the CB model (median RMSE>0.1). For each of these receptors, we searched as follows for additional interactions between receptors and odorants. If the observed receptor responses were higher than the predictions of the CB model, we hypothesized a synergistic interaction. If the observed receptor responses were lower than the CB model, we inferred the presence of suppression. We also looked at the composition of the mixtures for which the deviations were significant, and identified the common odorant (if any) and incorporated an interaction with this odorant compensating for over- or under- predictions. The parameters of the extended CB model were chosen, similar to the CB model, by minimizing the root mean squared error between observed response and predictions of the modified model weighted by the standard deviation. Applying this procedure to the the three remaining receptors significantly improved predictions (Fig. 4). Two receptors required inclusion of facilitative interactions (OR5P3, synergy between coumarin and acetophenone; Olfr1062, synergy between all three pairs), and one receptor (Olfr1104) required inclusion of suppression by eugenol (for functional forms and model parameters see Methods and SI Appendix, Modified models). Overall, the extended CB model (RMSE mean = 0.10, median = 0.06) outperformed a summation model (RMSE mean=0.17, median = 0.16) and shuffled model (RMSE mean=0.90, median = 0.86). These results

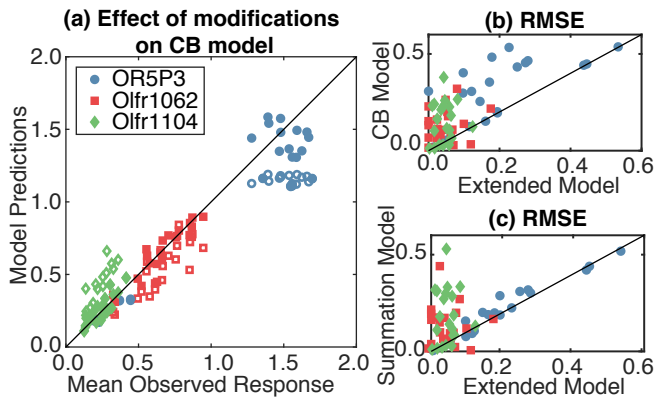

**Fig. 4. Adding synergy and suppression improves predictions for three receptors:** (a) The response of three olfactory receptors with CB model (open markers) and extended CB model (filled markers). Extensions: (i) OR5P3 – synergy between coumarin and acetophenone; (ii) Olfr1062 – synergy between all three pairs; (iii) Olfr1104 – suppression by eugenol. (b) RMSE of the CB model vs the extended CB model for the three receptors for each mixture. (c) RMSE of the summation model vs the extended CB model. In all panels, the diagonal line in black is the unit slope line.

predict specific odor-receptor interactions that can be tested experimentally.

## Discussion

In this work, we showed that a minimal biophysical model of odorant-receptor interaction incorporating just the simplest possible nonlinearity, namely competition between molecules for the binding site, can successfully predict the responses of many mammalian odor receptors to complex molecular mixtures. This is surprising because non-competitive interactions are common in pharmacology, but we nevertheless found that our simple model explains the majority of the experimental results. More general interactions between odorants and receptors can be easily added to our model, at the cost of additional parameters. For example, we showed that the nonlinearities implied by just competitive exclusion and facilitation are sufficient to produce diverse effects that have been previously reported in the perception of odor mixtures including synergy (17), overshadowing (16), suppression (31) and inhibition (29). These effects were thought to have a neural origin, but our results suggest that they may be driven partly by the biophysics of receptors.

Experimental studies of olfaction have largely focused on simple odors consisting of only one or two odorant molecules. However, natural odors are generally complex, containing hundreds of volatile components, with 3-40 being essential for the characteristic odor (28). Thus, in order to understand how olfactory circuits operate in naturalistic environments, models must account for complex sensory stimuli, as visual neuroscience has done for some time. A first step towards this goal is to understand how the receptors themselves respond to mixtures of many molecules. In practical terms, the combinatorial explosion of the number of mixtures with different compositions means that the only hope for progress is to have a model that can predict mixture responses from dose-response curves, which can conceivably be measured for large panels of odorants in high throughput experiments. Such a predictive model is most likely to be successful if it is rooted in the basic biophysics and biochemistry of molecular sensing, as our model is.

In olfaction, the low background activity of most receptors also makes it difficult to identify inverse agonists or antagonists using single molecules. But these effects, and more general non-competitive interactions, do occur in mixtures. Fortunately, such interactions will typically involve small numbers of molecules as the probability of multiple molecules meeting to interact at the same time should decline exponentially with the number of interacting molecules. Thus, future studies should be able to explore the landscape of interactions by testing receptor responses to mixtures with just a small number of components.

We demonstrated a strategy to identify such interactions and used it to identify some receptors with suppressive and synergistic interactions. Note that this process of identifying interactions will converge efficiently only if we begin at a biophysically well-motivated starting point like our competitive binding model. If we begin instead with an ad-hoc model like linear addition of responses, many corrections will be needed to get a good description, as in the accumulation of epicycles required to describe simple elliptical orbits in the Ptolemaic model of the solar system. Even if we start with the competitive binding model, the complexity of the added interactions must be discounted against the gain in accuracy, especially when including multiple interactions. This can be achieved via modern techniques in parametric statistical inference, e.g. (32), that trade off model complexity against prediction accuracy.

In the study of color vision, models of the early visual system are combined with look up tables of human responses to primary colors obtained through psychophysical experiments (33) to predict responses to arbitrary colors. These models have led to accepted industry standards that are used to produce color graphics through electronic or print means. Perhaps lookup tables of dose-response curves for olfactory receptors could be combined with models such as ours to predict responses to complex mixtures, ultimately allowing olfactory designers to create desired odors from a set of primary odorants.

**Materials and methods**

See SI Appendix for detailed Methods, biophysical models, and mathematical derivations.

**Measurement of dose-response curves and mixture response.** Receptor responses were measured as luminescence of Firefly and *Renilla* reporters in a cell based assay following the protocol for the Dual-Glo Luciferase Assay System (Promega) described in (24, 34). The enzyme is linear over seven orders of magnitude (35). In our system, luminescence from the firefly luciferase is a measure of receptor activity while luminescence from the *Renilla* luciferase measures how many cells are alive and successfully transfected. To measure receptor response, we first calculate the ratio of Firefly to *Renilla* luminescence on stimulation by the odor (see SI Appendix, Cell Based Assay). To standardize these measurements, we also measure the Firefly to *Renilla* luminescence ratio of a standard receptor (Olfr544) stimulated with nonanedioic acid at two concentrations (0 $\mu$M and 100 $\mu$M) under identical conditions. The luminescence ratio of the receptor is then divided by the difference between the luminescence ratio of the standard receptor at the two concentrations (see SI Appendix, Preprocessing). This gives the standardized response of a receptor to the odor. Lastly, we subtract the standardized response of the receptor at zero concentration of the odor to get the net response above baseline.

From 22 human and mouse receptors in (25), we selected 18 responding to at least two of eugenol, acetophenone, and coumarin (Sigma-Aldrich). We measured the dose-response curves to these odorants at 7 concentrations as well as a no-odor control. These 7 concentrations spanned the total concentration range allowed in our assay (up to 0.3 mM), which is much higher than the biologically-relevant concentrations found in the mucosa. We set a threshold for consistency that the difference between the standardized baseline response for a receptor to any pair of odorants should be within 0.2 of each other (see, e.g., the nearly overlapping baselines in Fig. 1a and Fig. 2a where this difference is nearly zero). 15 of the 18 receptors passed this test, and were further stimulated with 21 mixtures (12 binary, 9 ternary) of eugenol, coumarin and acetophenone (Methods; SI Appendix, Table 1) with concentrations selected to avoid receptor saturation.

From the data in (25), we also identified one receptor, Olfr168, that was broadly tuned, and for which dose-response curves were available for 12 odorants. We measured responses of this receptor to 24 mixtures of the 12 odorants and a no-odor control. Six mixtures contained all 12 odorants at equimolar concentrations. To select the other 18 mixtures, we first fit our competitive binding model to the dose-response data and used it to select pseudo-random concentrations of each odorant such that the predicted responses spanned the full dynamic range while avoided saturation (compositions in SI Appendix, Table 3).

**Model parameter estimation using dose-response measurements.** For each odorant ($i$), we chose parameters ($EC50_i$ and the product $F_{\max}e_i$) that minimize the root mean squared error between the measured average response ($\bar{y}_{ex}(c_i)$) at concentrations $c_i$ and the model predictions ($F(c_i)$), divided by the experimental standard deviation (details in SI Appendix, Model parameter estimation), i.e.,

$$E_i = \sqrt{\frac{1}{M} \sum_{c_i} \left( \frac{(F(c_i) - \bar{y}_{ex}(c_i))}{\sigma(c_i)} \right)^2} \qquad [2]$$

$E_i < 1$ would mean that, on average, the model predictions lie within one standard deviation away from the mean experimental observation. The minimization was performed using MATLAB *fminunc*. (Also see SI Appendix, Dealing with unconstrained parameters and SI Appendix, Fig. S5 and S6, for an alternative procedure for parameter estimation).

**Null models.** We considered a *Summation Model* where the receptor response to mixtures was a sum of the response to individual odorants at their concentrations in the mixture (see SI Appendix, Summation Model). We also considered a *Shuffled Model* that has the same mathematical form as the competitive binding model (Eq. 1), but with parameters chosen randomly with replacement from the set of dose-response parameters used in our analysis (57 sets; 45 sets from the 15 receptors of the binary-ternary analysis and 12 sets of the receptor Olfr168 from the 12-component analysis). Each parameter of the shuffled model is chosen independently. We report average prediction error (RMSE) over 300 such random choices.

**Competitive binding model and extensions.** Mathematical derivation of the models from the biophysics of molecular binding is given in SI Appendix. The model for synergistic interaction (see SI Appendix, Facilitation) has the form:

$$F(c_1, c_2) = \frac{F_{max}\left(e_1 \frac{c_1}{EC50_1} + e_2 \frac{c_2}{EC50_2} + e_{12} \frac{c_1 c_2}{EC50_{12}}\right)}{\left(1 + \frac{c_1}{EC50_1} + \frac{c_2}{EC50_2} + \frac{c_1 c_2}{EC50_{12}}\right)} \quad [3]$$

where $e_{12}$ and $EC50_{12}$ are the parameters of the interaction between the two odorants. The model with suppression (see SI Appendix, Non-competitive inhibition) has the form:

$$F(c_1, c_2) = \frac{F_{max}\left(e_1 \frac{c_1}{EC50_1} + e_2 \frac{c_2}{EC50_2}\right)}{\left[1 + \frac{c_1}{EC50_1} + \left(\frac{c_2}{EC50_2}\right)\left(1 + K_1 \frac{c_1}{EC50_1}\right)\right]} \quad [4]$$

where $K_1$ is the suppression parameter for odor 1.

**Data and Software Availability.** Data and software are available upon request from the authors.

**ACKNOWLEDGMENTS.** VB and VS were supported by the Simons Foundation (Grant 400425; Mathematical Modeling of Living Systems). VB was supported by US-Israel Binational Science Foundation grant 2011058 and Physics Frontiers Center grant PHY-1734030. Work by VB at the Aspen Center for Physics was supported by NSF grant PHY-1607611. NM and JM were supported by R01 DC013339. A portion of the work was performed using the Monell Chemosensory Receptor Signaling Core and Genotyping and DNA/RNA Analysis Core, which was supported, in part, by funding from the US National Institutes of Health NIDCD Core Grant P30 DC011735.

## Supporting Information Text

## Competitive binding model

Consider a receptor $R$ interacting with a mixture of two odorants ($O_1$ and $O_2$) with concentrations $c_1$ and $c_2$, and binding affinities $k_1$ and $k_2$ respectively. Assume that only one odorant molecule can bind to a receptor binding site at a time. Upon binding, the odorant molecule forms a receptor-odor complex $O_iR$, ($i = 1, 2$). The complex can either dissociate with rate $r_i$, giving back the receptor and the odorant molecule, or can enter an active state $O_iR^*$ with rate $k_i^*$ which elicits a detectable response. In our experimental assay, not every odorant binding event would lead to luminescence. The active state receptor represents those receptors that produce luminescence. In nature, these are the receptors that result in a downstream response after binding.

The complex in the active state can revert back to the inactive state with rate $r_i^*$. The total response is proportional to the total number of receptor-odor complexes in the active state. These interactions can be summarized by the following chemical reactions:

$$O_1 + R \underset{r_1}{\overset{k_1}{\rightleftharpoons}} O_1R \underset{r_1^*}{\overset{k_1^*}{\rightleftharpoons}} O_1R^* \qquad [1a]$$

$$O_2 + R \underset{r_2}{\overset{k_2}{\rightleftharpoons}} O_2R \underset{r_2^*}{\overset{k_2^*}{\rightleftharpoons}} O_2R^* \qquad [1b]$$

These reactions can be described by the following equations

$$\frac{dR}{dt} = -(k_1 c_1 R + k_2 c_2 R) + (r_1 R_1 + r_2 R_2) \qquad [2a]$$

$$\frac{dR_i}{dt} = (k_i c_i R + r_i^* R_i^*) - (r_i R_i + k_i^* R_i) \qquad [2b]$$

$$\frac{dR_i^*}{dt} = k_i^* R_i - r_i^* R_i^* \qquad [2c]$$

where i=1,2, with $R$, $R_i$, $R_i^*$ being the concentrations of receptors that are unbound, bound to odorant $i$ but inactive, and bound to odorant $i$ in an active state, respectively. The sum of all these concentrations is fixed to be $R_{max}$ reflecting the total number of available receptors. Assuming that the response is proportional to the total number of bound receptors in the active state ($R^* = R_1^* + R_2^*$), we can solve the rate equations at steady state to find that

$$F(c_1, c_2) = \frac{F_{max}\left(e_1 \frac{c_1}{EC50_1} + e_2 \frac{c_2}{EC50_2}\right)}{\left(1 + \frac{c_1}{EC50_1} + \frac{c_2}{EC50_2}\right)} \qquad [3]$$

where $EC50_i = r_i r_i^*/(k_i(k_i^* + r_i^*))$ is the concentration at which the response is half of the maximum for odorant $i$, $e_i = k_i^*/(k_i^* + r_i^*)$ is the efficacy of the receptor for the odorant $i$, and $F_{max}$ is a parameter that depends on the total receptor concentration. Fig. 3 (main text) shows the response of a receptor according to this model. For an N component mixture with odorant concentrations $\{c_i : i = [1, N]\}$, this result generalizes to:

$$F(\{c_i\}) = \frac{F_{max} \sum_{i=1}^{N}\left(e_i \frac{c_i}{EC50_i}\right)}{\left(1 + \sum_{i=1}^{N} \frac{c_i}{EC50_i}\right)} \qquad [4]$$

## Extension of the competitive binding model

**Facilitation.** It is possible that the binding of one odorant facilitates the binding of other odorants. In a mixture, such interactions can be considered through the following chemical reaction:

$$O_1 + R \underset{r_1}{\overset{k_1}{\rightleftharpoons}} O_1R \underset{r_1^*}{\overset{k_1^*}{\rightleftharpoons}} O_1R^* \qquad [5]$$

$$O_2 + R \underset{r_2}{\overset{k_2}{\rightleftharpoons}} O_2R \underset{r_2^*}{\overset{k_2^*}{\rightleftharpoons}} O_2R^* \qquad [6]$$

$$O_1R + O_2 \underset{r_{12}}{\overset{k_{12}}{\rightleftharpoons}} O_1O_2R \underset{r_{12}^*}{\overset{k_{12}^*}{\rightleftharpoons}} O_1O_2R^* \qquad [7]$$

The solution of the corresponding rate equations at steady state is

$$F(c_1, c_2) = \frac{F_{max}\left(e_1 \frac{c_1}{EC50_1} + e_2 \frac{c_2}{EC50_2} + e_{12} \frac{c_1 c_2}{EC50_{112}}\right)}{\left(1 + \frac{c_1}{EC50_1} + \frac{c_2}{EC50_2} + \frac{c_1 c_2}{EC50_{112}}\right)} \qquad [8]$$

where $EC50_i = r_i r_i^*/(k_i(k_i^* + r_i^*))$, $e_i = k_i^*/(k_i^* + r_i^*)$, $EC50_{112} = r_1 r_{12} r_{12}^*/(k_1 k_{12}(k_{12}^* + r_{12}^*))$ and $e_{12} = k_{12}^*/(k_{12}^* + r_{12}^*)$. This simple modification with different parameters sets $p = \{EC50_1, e_1, EC50_2, e_2, EC50_{112}, e_{12}, c_1/c_2\}$ can account for behaviors like synergy (Fig. 4d; $p = \{10^{-5.5}, 0.4, 10^{-5}, 0.3, 10^{-11}, 0.85, 1\}$) in which the receptor response to mixture is higher than all constituent odorants, overshadowing ($p = \{10^{-7}, 0.9, 10^{-5}, 0.4, 10^{-3}, 0.3, 1\}$) where the mixture response is dominated by the odorant with highest individual response, suppression ($p = \{10^{-5}, 0.9, 10^{-5}, 0.4, 10^{-3}, 0.5, 1\}$) where the mixture response is in between the individual response of the odorants, overshadowing ($p = \{10^{-4}, 0.9, 10^{-6}.5, 0.5, 10^{-3}, 0.3, 1\}$) where the mixture response is dominated by the odorant with lowest individual response, and inhibition (Fig. 4e; $p = \{10^{-5}, 1, 10^{-5.5}, 0.6, 10^{-12}, 0.35, 1\}$) in which the response is lower than the responses to all individual odorants.

**Independent binding sites.** If there are independent binding sites on the receptor molecule for different odorants, then the receptor response to the mixture would simply be the sum of the response of individual odorants. For a binary mixture, the response would be given as:

$$F(c_1, c_2) = \frac{F_{max}\left(e_1 \frac{c_1}{EC50_1}\right)}{\left(1 + \frac{c_1}{EC50_1}\right)} + \frac{F_{max}\left(e_2 \frac{c_2}{EC50_2}\right)}{\left(1 + \frac{c_2}{EC50_2}\right)} \quad [9]$$

**Non-competitive inhibition.** If the odorants bind independently to two binding sites, but the receptor responds when only the first binding site is occupied, we can have the following set of reactions:

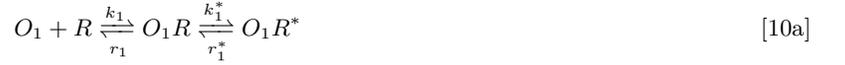
$$O_1 + R \underset{r_1}{\overset{k_1}{\rightleftharpoons}} O_1 R \underset{r_1^*}{\overset{k_1^*}{\rightleftharpoons}} O_1 R^* \quad [10a]$$

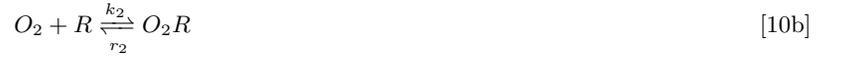
$$O_2 + R \underset{r_2}{\overset{k_2}{\rightleftharpoons}} O_2 R \quad [10b]$$

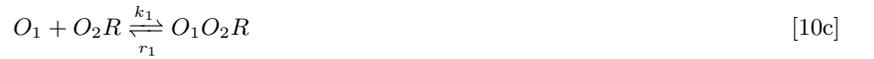
$$O_1 + O_2 R \underset{r_1}{\overset{k_1}{\rightleftharpoons}} O_1 O_2 R \quad [10c]$$

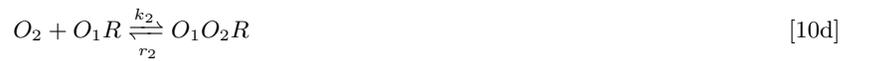
$$O_2 + O_1 R \underset{r_2}{\overset{k_2}{\rightleftharpoons}} O_1 O_2 R \quad [10d]$$

These set of equations allows for 5 possible states of the receptor. Unbound ($R$), bound with odorant 1 or 2 ($O_1 R, O_2 R$), bound with both odorants ($O_1 O_2 R$) and in the bound excited state ($O_1 R^*$). The receptor response is proportional to the concentration of $O_1 R^*$ and is given by:

$$F(c_1, c_2) = \frac{F_{max}\left(e_1 \frac{c_1}{EC50_1}\right)}{\left(1 + \frac{c_1}{EC50_1} + K_2 c_2(1 + K_1 c_1)\right)} \quad [11]$$

where $K_i^* = k_i/r_i$. Such interactions lead to suppression and inhibition.

**Independent binding sites with facilitation.** The binding of one odorant on a receptor binding site may facilitate the binding of the other odorant on the second binding site. Such an interaction leads to the following chemical reactions:

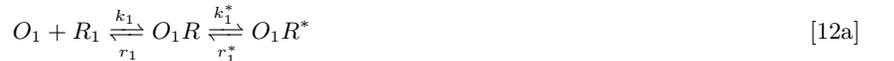
$$O_1 + R_1 \underset{r_1}{\overset{k_1}{\rightleftharpoons}} O_1 R \underset{r_1^*}{\overset{k_1^*}{\rightleftharpoons}} O_1 R^* \quad [12a]$$

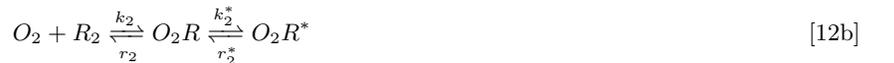
$$O_2 + R_2 \underset{r_2}{\overset{k_2}{\rightleftharpoons}} O_2 R \underset{r_2^*}{\overset{k_2^*}{\rightleftharpoons}} O_2 R^* \quad [12b]$$

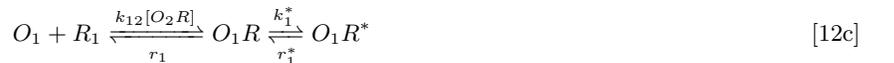
$$O_1 + R_1 \underset{r_1}{\overset{k_{12}[O_2 R]}{\rightleftharpoons}} O_1 R \underset{r_1^*}{\overset{k_1^*}{\rightleftharpoons}} O_1 R^* \quad [12c]$$

These equations can be solved as earlier under the constraints $(R_1 + O_1 R + O_1 R^* = F_{max})$ and $(R_2 + O_2 R + O_2 R^* = F_{max})$. The receptor response is given as:

$$F(c_1, c_2) = \frac{F_{max}}{\left(\frac{1}{e_1} + \frac{\left(1 + \frac{c_2}{EC50_2}\right)}{\left(e_1 \frac{c_1}{EC50_1}\left(1 + e_2 \frac{c_2}{EC50_2}\left(\frac{F_{max} k_{12} r_2^*}{k_1 k_2^*} + \frac{1}{e_2}\right)\right)\right)}\right)} + F_{max} \frac{e_2 \frac{c_2}{EC50_2}}{\left(1 + \frac{c_2}{EC50_2}\right)} \quad [13]$$

This interaction results in synergy and overshadowing by the more reactive odor.

**Odorant dimerization.** To include cases where different odorant molecules combine to form a heterodimer ($O_1O_2$) before binding to the receptor, we can consider the following chemical reaction:

$$O_1 + R \underset{r_1}{\overset{k_1}{\rightleftharpoons}} O_1R \underset{r_1^*}{\overset{k_1^*}{\rightleftharpoons}} O_1R^* \tag{14a}$$

$$O_2 + R \underset{r_2}{\overset{k_2}{\rightleftharpoons}} O_2R \underset{r_2^*}{\overset{k_2^*}{\rightleftharpoons}} O_2R^* \tag{14b}$$

$$O_1 + O_2 \underset{a_2}{\overset{a_1}{\rightleftharpoons}} O_1O_2 \tag{14c}$$

$$O_1O_2 + R \underset{r_{12}}{\overset{k_{12}}{\rightleftharpoons}} O_1O_2R \underset{r_{12}^*}{\overset{k_{12}^*}{\rightleftharpoons}} O_1O_2R^* \tag{14d}$$

The solution of the corresponding rate equations at steady state is

$$F(c_1, c_2) = \frac{F_{max}\left(e_1 \frac{c_1}{EC50_1} + e_2 \frac{c_2}{EC50_2} + e_{12} \frac{ac_1c_2}{EC50_{12}}\right)}{\left(1 + \frac{c_1}{EC50_1} + \frac{c_2}{EC50_2} + \frac{ac_1c_2}{EC50_{12}}\right)} \tag{15}$$

Here $a = a_1/a_2$ is the ratio of the binding and unbinding rates of the odorant molecules. This equation is similar to odorant facilitation. The resulting effects are synergy, overshadowing, suppression and inhibition.

## Materials and methods

**Cell-based assay.** In vitro experiments followed the protocol for a Dual-Glo Luciferase Assay System (Promega) described in (1, 2). Hana3A cells (courtesy of Matsunami Laboratory), were plated into 96-well poly-D-lysine-coated plates (Corning BioCoat). Negative mycoplasma status and cell line identity were confirmed for cells used in these experiments (ATCC; Promega). Plated cells were transfected with 5 ng/well of RTP1S-pCI (1, 3), 5 ng/well of pSV40-RL, 10 ng/well pCRE-luc, 2.5 ng/well of M3-R-pCI (4), and 5 ng/well of plasmids containing rhodopsin-tagged olfactory receptors. Each plate was transfected with eight wells of Olfr544 as a standard. Twenty-four hours after transfection, we applied each monomolecular odorant or mixture in quadruplicate. Odors were diluted to the final concentration in CD293 (ThermoFisher Scientific). To standardize across plates, half of the standard wells were stimulated with CD293 (no odor) and half were stimulated with $100\mu$M Nonanedioic acid (Sigma-Aldrich) in CD293. All olfactory receptors on a given plate were stimulated with the same odorant or mixture. Four hours after odor stimulation, we measured luminescence according to the Dual-Glo protocol (BioTek Synergy 2 reader).

**Binary and ternary mixtures.** Based on data from (5), we identified 22 receptors likely to respond to eugenol, acetophenone, and coumarin (Sigma-Aldrich). We applied seven steps of three-fold serial dilutions for each odorant starting at 0.3 mM as well as a no-odor control. Odors were applied in quadruplicate to all 22 receptors to obtain dose response curves for each receptor. Four receptors, Olfr362, Olfr315, Olfr1110, Olfr165, were not tested with subsequent mixtures because they did not respond to at least two of the three odorants. We set a threshold for consistency that the standardized baseline responses for the same receptor to different odors should be within 0.2 of each other – three receptors, OR2W1, Olfr558 and Olfr620, were removed from the analysis by this criterion. For the 15 remaining receptors, we measured the responses to twelve binary and nine ternary mixtures of eugenol, acetophenone, and coumarin (see Table S1 for mixture compositions).

**12-component mixtures.** From previously collected measurements of olfactory receptor responses to monomolecular odorants (6) we identified eight receptors that are broadly tuned. Of these we selected receptor Olfr168 (which had dose-response data for 23 odorants) and chose the twelve odorants with the lowest EC50s for further experiments (list in Fig 2a, inset). We first used the dose response data from (6) to estimate competitive-binding model parameters. We then tested 24 mixtures of 12 odorants as well as a no-odor control. Six of the mixtures contained all 12 odorants at equimolar concentrations. For the other 18 mixtures, we used the model to select pseudo-random concentrations of each odorant such that the predicted receptor response spanned the full dynamic range and avoided saturation (see Table S3 for the mixture compositions).

**Odorants.** Odorants (Sigma-Aldrich; See Table S2) were diluted to the final concentration in CD293 (ThermoFisher Scientific) with the exception of four odorants (4-Chromanone, acetophenone, coumarin, and eugenol) that were diluted from 1M stocks in DMSO. All mixtures containing odorants diluted from 1M stocks in DMSO had less than 0.05% DMSO in the final mixture.

## Quantification and statistical analysis.

***Preprocessing.*** We divided the Firefly luciferase luminescence by *Renilla* luciferase luminescence to normalize for transfection efficiency and cell death. In our system, Firefly is a measure of receptor activity while *Renilla* is made regardless of the receptor activity and is a measure of the response capacity, i.e., how many cells are alive and successfully transfected. Thus, we are normalizing the response capacity of the cells in the specific assay. Additionally, we divided by the difference between the standard Olfr544 response at $0\mu M$ and $100\mu M$ Nonanedioic acid. This preprocessing allowed us to compare receptor responses relative to our standard receptor and to standardize responses across plates. Lastly, we subtracted the standardized receptor response at zero concentration of odorant/mixture to get the net response above baseline: $\left(\frac{r}{(s_{100}-s_0)} - \frac{r_0}{(s_{100}-s_0)}\right)$.

***Model parameter estimation.*** First, we used the experimentally observed response ($y_{ex}(c_i)$) of each receptor to odorant $i$ at concentration $c_i$ measured over four replicates to calculate the average receptor response ($\bar{y}_{ex}(c_i)$) and the standard deviation ($\sigma(c_i)$). The individual dose response was measured at $M$ concentrations, where $M = 8$ for binary-ternary analysis and $M = 11$ for 12-component analysis. For each odorant, we chose parameters ($EC50_i$ and the product $F_{\max}e_i$) to minimize the mean squared error between this average measured response and the model prediction ($F(c_i)$), weighted by the experimental standard deviation, i.e.:

$$E_i = \sqrt{\frac{1}{M}\sum_{c_i}\left(\frac{(F(c_i)-\bar{y}_{ex}(c_i))}{\sigma(c_i)}\right)^2} \quad [16]$$

$E_i = 1$ would mean that, on average, the model predictions lie one standard deviation away from the mean experimental observation. As $E_i$ is a function of the $EC50_i$ and the products $F_{\max}e_i$, we choose these parameter combinations to minimize $E_i$. See Table S4 for DR parameters for the binary-ternary mixture analysis and Table S5 for 12-component mixture analysis.

***Standard deviation in CB model predictions.*** The standard deviations in the prediction of the CB model were estimated as the standard deviation over CB model predictions of 300 randomly chosen sets of model parameters that allow dose-response predictions such that Eq.16 was lower than $\lceil E_i^{min}\rceil$. $E_i^{min}$ is the error for the best fit parameters. $\lceil x \rceil$ is the ceiling operation which represents the smallest integer greater than $x$. Thus, if $E_i^{min} < 1$ the parameters were chosen such that the DR predictions were at most 1 standard deviation from the mean observations, if $1 < E_i^{min} < 2$, the parameters were chosen such that the DR predictions were at most 2 standard deviation from the mean observations, etc.

***Quantifying model performance.*** We quantified the performance of the model in terms of the root mean squared error (RMSE) defined as:

$$RMSE = \sqrt{\frac{1}{N_m}\sum_{i=1}^{N_m}(F_i-y_i)^2}, \quad [17]$$

where $F_i$ is the model prediction for a mixture, $y_i$ is the corresponding mean experimental observation and $N_m$ is the number of mixtures over which the error is calculated.

***Additional measures of prediction error.*** In addition to RMSE, we also looked at other measures of prediction error (Figure S1). For every mixture, we compared the RMSE to the observed receptor response (relative error, Figure S1a), and found that the prediction error is typically small compared to the response magnitude of each receptor to each mixture. To compare the CB model to the typical receptor response, we compared the RMSE to mean response over the mixtures used in the experiment (Figure S1b), and found that the prediction error is small compared to the response range of each receptor. Next, we showed that the RMSE is comparable to the standard deviation in experimental measurements (Figure S1c). Lastly, we looked at the robustness of the CB model, by comparing the standard deviation of the CB model predictions to the experimental standard deviation (see above for the procedure used to define the model standard deviation; results in Figure S1c). The CB model standard deviations are small compared to the experimental standard deviation, confirming robustness of the model.

**Alternative models for comparison.** 1. Summation Model: The receptor response in the summation model is given by

$$F_{\text{summation}}(c_i) = \sum_{i=1}^{N}\frac{F_{\max}\left(\frac{e_ic_i}{EC50_i}\right)}{\left(1+\frac{c_i}{EC50_i}\right)} \quad [18]$$

where $c_i$'s are the concentrations of odorants.

2. Shuffled Model: The odor shuffled model has the same mathematical form as Eq. 1 in the main text, but parameters are shuffled by choosing each parameter value randomly from the dose-response sets used in our analysis. For example, the EC50 for one receptor would be assigned one of the EC50 values from a set of 57 values (45 values from the dose-responses of binary-ternary mixture components and 12 values from the dose-responses of the 12-component mixture components). These include the 45 sets of the binary-ternary analysis and 12 sets of the 12-component analysis, for a total of 57 sets of parameters. We estimated the response over 300 random shuffles of these parameters.

**Modified models.**

1. OR5P3: (Synergy) For OR5P3 we included a synergistic interaction between acetophenone and coumarin. The mixture response was given as:

$$F(c_1, c_2, c_3) = \frac{F_{max}\left(e_1 \frac{c_1}{EC50_1} + e_2 \frac{c_2}{EC50_2} + e_3 \frac{c_3}{EC50_3} + e_{23} \frac{c_2 c_3}{EC50_{23}}\right)}{\left(1 + \frac{c_1}{EC50_1} + \frac{c_2}{EC50_2} + \frac{c_3}{EC50_3} + \frac{c_2 c_3}{EC50_{23}}\right)} \quad [19]$$

where the subscripts 1, 2 and 3 refers to odorants eugenol, acetophenone and coumarin respectively. The interaction parameters are $\{e_{23}, EC50_{23}\} = \{2.161, -9.000\}$. These parameters were obtained by minimizing the root mean squared error between observed response and predictions of the modified model weighted by the standard deviation. The other parameters are given in Table S4.

2. Olfr1062: (Synergy) For Olfr1062 synergistic interactions were included for all three pairs of receptors. The mixture response was given as:

$$F(c_1, c_2, c_3) = \frac{F_{max}\left(e_1 \frac{c_1}{EC50_1} + e_2 \frac{c_2}{EC50_2} + e_3 \frac{c_3}{EC50_3} + e_{12} \frac{c_1 c_2}{EC50_{12}} + e_{23} \frac{c_2 c_3}{EC50_{23}} + e_{31} \frac{c_3 c_1}{EC50_{31}}\right)}{\left(1 + \frac{c_1}{EC50_1} + \frac{c_2}{EC50_2} + \frac{c_3}{EC50_3} + \frac{c_1 c_2}{EC50_{12}} + \frac{c_2 c_3}{EC50_{23}} + \frac{c_3 c_1}{EC50_{31}}\right)} \quad [20]$$

where $\{e_{12}, EC50_{12}, e_{23}, EC50_{23}, e_{31}, EC50_{31}\} = \{0.4572, -8.4268, 1.3659, -7.9642, 1.1406, -8.0314\}$. These parameters were obtained by minimizing the root mean squared error between observed response and predictions of the modified model weighted by the standard deviation. The rest of the parameters are given in Table S4.

3. Olfr1104: (Suppression) For Olfr1104 we included suppression by eugenol. The mixture response was given as:

$$F(c_1, c_2, c_3) = \frac{F_{max}\left(e_1 \frac{c_1}{EC50_1} + e_2 \frac{c_2}{EC50_2} + e_3 \frac{c_3}{EC50_3}\right)}{\left[1 + \frac{c_1}{EC50_1} + \left(\frac{c_2}{EC50_2} + \frac{c_3}{EC50_3}\right)\left(1 + K_1 \frac{c_1}{EC50_1}\right)\right]} \quad [21]$$

where the subscripts 1, 2 and 3 refers to odorants eugenol, acetophenone and coumarin respectively and $K_1 = 42.87$. The suppression parameter was obtained by minimizing the root mean squared error between observed response and predictions of the modified model weighted by the standard deviation. The rest of the parameters are given in Table S4.

**Chi-squared test.** We estimated the chi-squared statistic ((observed − expected)$^2$/expected, where observed is the mean experimentally observed response and expected is the expected CB model prediction) to test the hypothesis that the receptor response is generated by the CB model. The chi-squared statistic for the entire dataset of 12 receptors responding to 21 binary-ternary mixtures was 4.55 (degree of freedom 251), with $p > 0.999$. The chi-squared statistics for the individual receptors were in the range 0.10-1.2, (degree of freedom 20), again with $p > 0.999$. For the 12 component mixture, the chi-squared statistic is 2.08 (degree of freedom 23), with $p > 0.999$. Thus, with high confidence, the hypothesis that the receptor response is generated by the CB model is not rejected.

**Dealing with unconstrained parameters.** The predictive power of our model depends in part on the quality of parameters inferred from responses to single odorants. For good estimates of the parameters, the experiment should capture three characteristic parts of the dose-response curve: the threshold at low concentrations, monotonic increase of response at intermediate concentrations, and saturation at high concentrations. In practice, our ability to measure response saturation of olfactory receptors was limited because cells in our preparation did not survive exposure to odorant concentrations higher than approximately 0.3 mM. In such cases, the model parameters are not fully constrained and there is a range of parameter choices that gives an equally good description of the measured responses to single odorants (SI Fig. S5). For example, for the receptor Olfr895 in the region where the cost function is within 5% of the minimum value, the value of logEC50 for eugenol varies by +/- 4% of the best value, for acetophenone the variation is +/- 12%, and for coumarin the variation is +/- 3%. Such unconstrained directions in parameter space are common in biological experiments (7, 8), often because of the difficulties of sampling the extremes of nonlinear response functions. This limitation cannot be overcome by collecting more data at lower concentrations. To test robustness of our predictions with respect to parameter variations we varied the parameters derived from dose-response curves as described above in the section entitled "Standard deviation in CB model predictions". The resulting variation in model predictions for mixture responses was typically well within the experimental measurement standard deviation (Fig S1d and SI section "Additional measures of prediction error").

As an alternative approach, we used a subset of the mixture response data to constrain the parameters. We formed a training set of receptor responses to dose-response measurements and a subset of mixture response measurements. We estimated parameters by combining the relative squared error for the odorant dose-response and mixture training set response as:

$$\sqrt{\frac{1}{N_o}\sum_{i=1}^{N_o} E_i^2 + \frac{1}{N}\sum_{k=1}^{N}\left(\frac{(F(m_k) - \bar{y}_{ex}(m_k))}{\sigma_{(m_k)}}\right)^2} \quad [22]$$

The first term under the square root is the mean of the squared error, weighted by the standard deviation, of all dose-response measurements, with $N_o$ being the total number of odorants for which the dose-response was measured ($N_o = 3$ for the binary/ternary experiments). The second term under the square root is the squared error, weighted by the standard deviation, for the mixture measurements in the training set. Here, $F(m_k)$ is the receptor response to mixture $m_k$, and $\bar{y}_{ex}(m_k)$ and $\sigma_{m_k}$ are the corresponding experimentally observed mean response and standard deviation calculated over 4 replicates of the mixture. $N$ is the number of mixtures in the training set. For the binary-ternary analysis, the mixture training set had $N = 3$, the minimum number of data points required to constrain three dose-response curves. For a fair comparison with the model trained only on dose-response data, we removed three measurements from the dose-response data, one for each odorant.

This procedure constrains the parameters to be consistent across the dose-response measurements and the mixture measurements as well. Notice that the minimization does not infer any mixture interaction, but instead provides a better estimate of the individual dose-response parameters. The parameters thus obtained were used to predict the response to the remaining mixtures. To avoid any overlap, the mixtures used for training and for prediction were chosen to have different compositions. For the 12 receptors used in binary/ternary analysis, we used only three binary mixtures to constrain the parameters, but both binary and ternary mixtures to evaluate the model.

Including mixture data does not affect the prediction quality for responses to single odorants, and modestly improves the predictions for held out binary-ternary mixtures (Fig. S6) for most receptors. Thus, while the CB model parameters can be estimated from dose-response response data, when experimental constraints limit dose-response measurements the response to small (binary) mixtures can be used to supplement dose-response measurements for model parameter estimation.

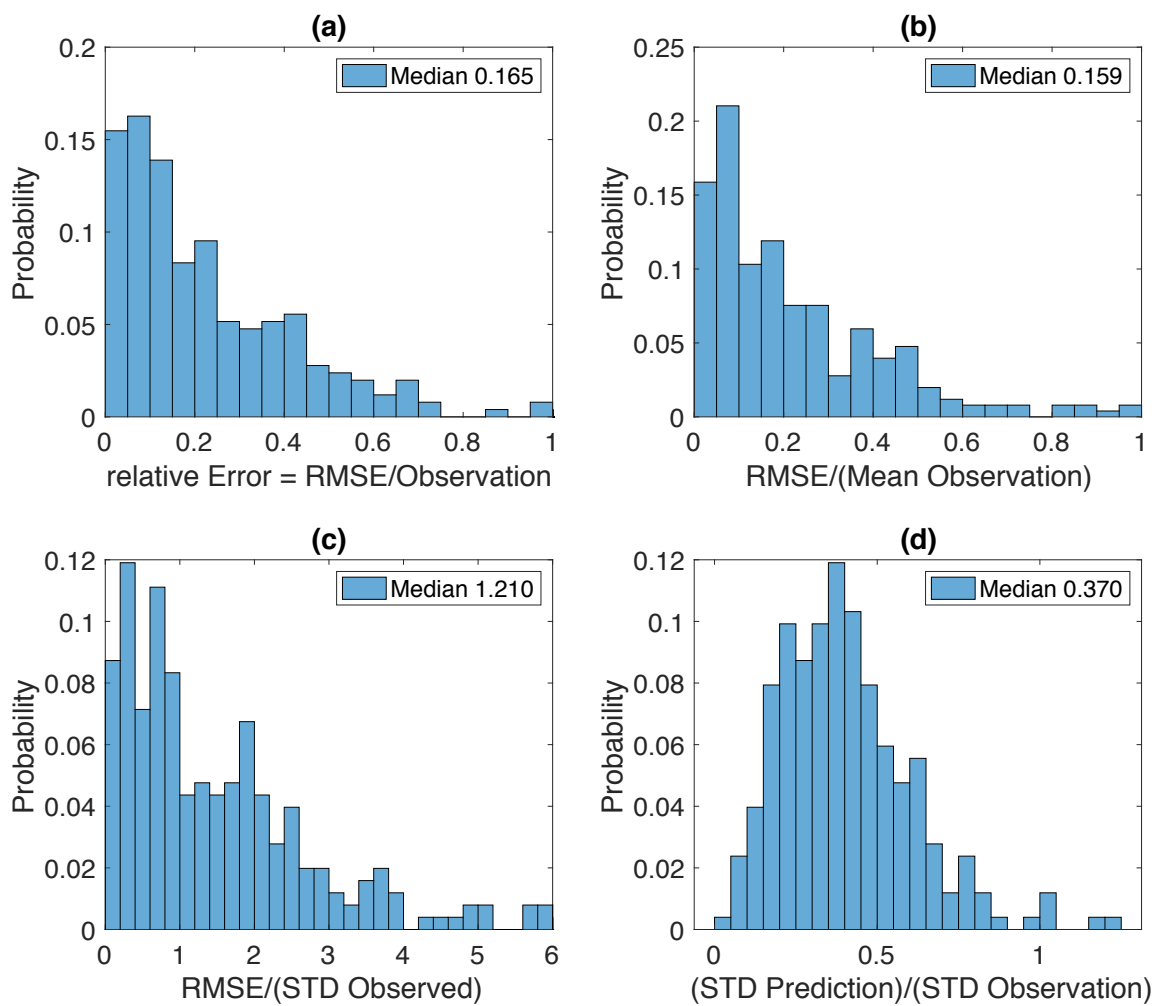

**Fig. S1.** Binary Ternary Error Comparison: (a) Histogram of the ratio of RMSE to the observed response for each mixture. The data has been pooled over all 12 receptors shown in figure 1 in the main text. The low median shows that the prediction error is typically small compared to the response magnitude of each receptor to each mixture. (b) Histogram of the ratio of RMSE to the mean observed response of the receptors. The small median shows that the prediction error is small compared to the response range of each receptor. (c) Histogram of the ratio of RMSE to the experimentally observed standard deviation. A ratio lower than 1 indicates that the predictions are within one standard deviation from the observed response. (d) Histogram of the ratio of standard deviation of CB model predictions and experimental observations. As described in the SI Text the model standard deviation is computed by allowing the model parameters to vary such that the dose-curves remain within one standard deviation of the best fit. The variation in the theory predictions for mixture responses is within the experimental standard deviation.

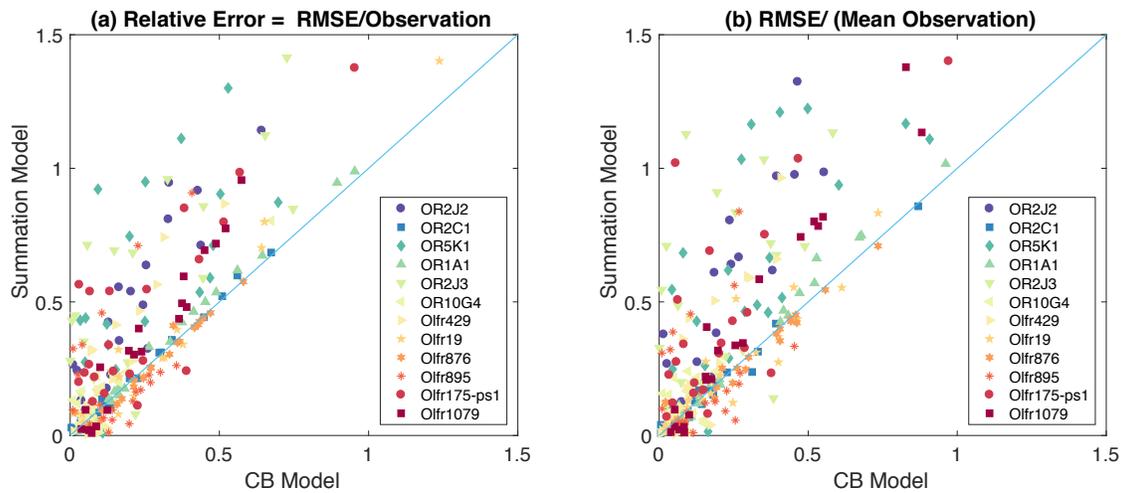

**Fig. S2.** Model comparison using other error metrics: (a) Plot of relative error of summation model v/s the CB model. The relative error is calculated as the RMSE divided by the experimental observation. The relative errors of the summation model are above the unit slope line, indicating that the CB model predictions outperform the summation model predictions. (b) Plot of RMSE divided by the mean observed response of the receptor. For each receptor the mean observed response is calculated over all mixtures used in the binary-ternary experiment. The CB model predictions are better than the summation model.

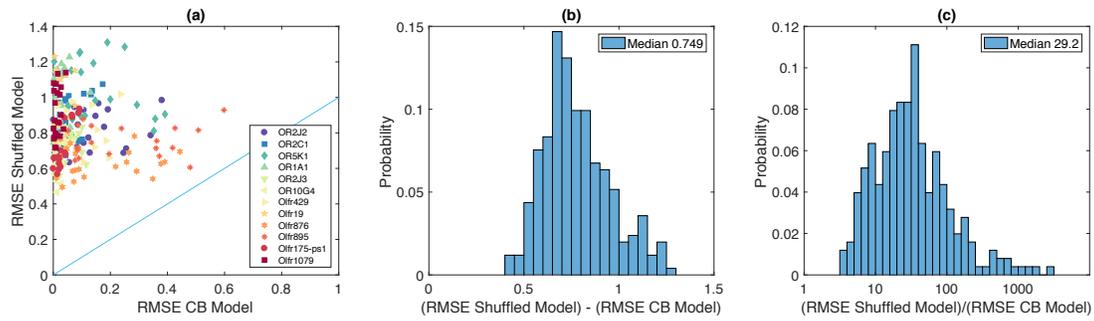

**Fig. S3.** Comparison of CB model to shuffled model for the 12 receptors in Figure 1 of main text. The shuffled model has the same functional form as the CB model, but the parameters were chosen randomly from all the dose-response parameters available in our set. (a) The RMSE of the shuffled model plotted vs the RMSE of the CB model. The CB model outperforms the shuffled model. (b) Histogram of difference between the RMSE of shuffled and CB model. The difference is dominated by the RMSE of the shuffled model. (c) The histogram of the ratio between RMSE of shuffled model and CB model. The RMSE of shuffled model are 10-100 times larger than the CB model.

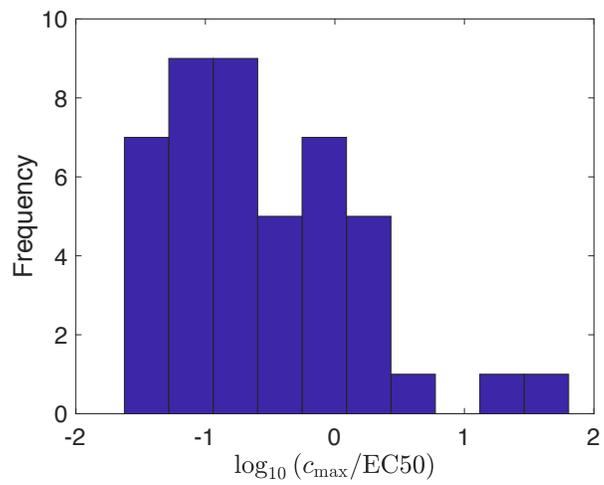

**Fig. S4.** Maximum tested concentrations relative to the receptor EC50 for the 15 receptors (15*3=45 EC50 values) used in binary/ternary mixtures, each receptor tested against eugenol, coumarin, and acetophenone ($c_{\max}$ =0.3 mM). Due to cell death at high odorant concentrations, it is not possible to probe the saturation regime ($\log_{10}(c_{\max}/\text{EC50}) \gg 0$) for all receptor-odor pairs.

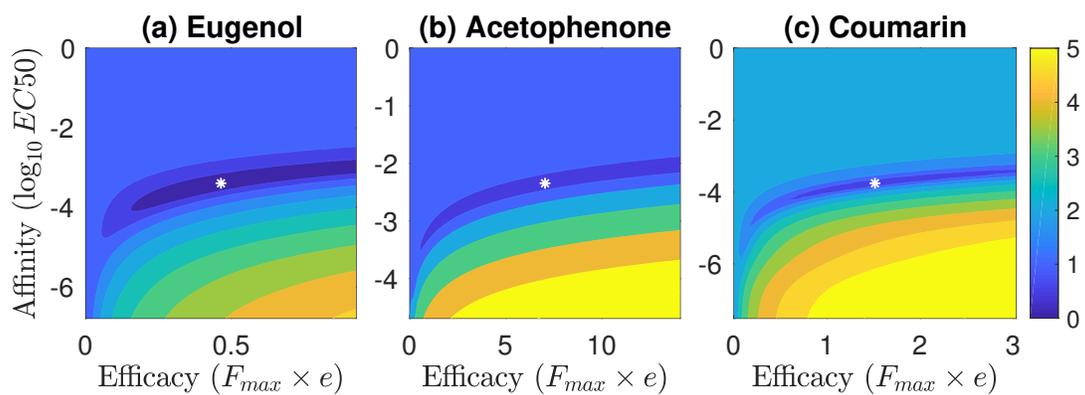

**Fig. S5.** The fitting error ($log_{10} E_i$ from Eq. 16) varies with the affinity and efficacy parameters, shown here for receptor Olfr895 responding to Eugenol, Coumarin, and Acetophenone. The white asterisks label the minimum $E_i$. There is a narrow region (deep blue) where the cost function $E_i$ does not change appreciably while $EC50$ and $F_{max} \times e$ are varied, showing that additional information is necessary to fully constrain these parameters.

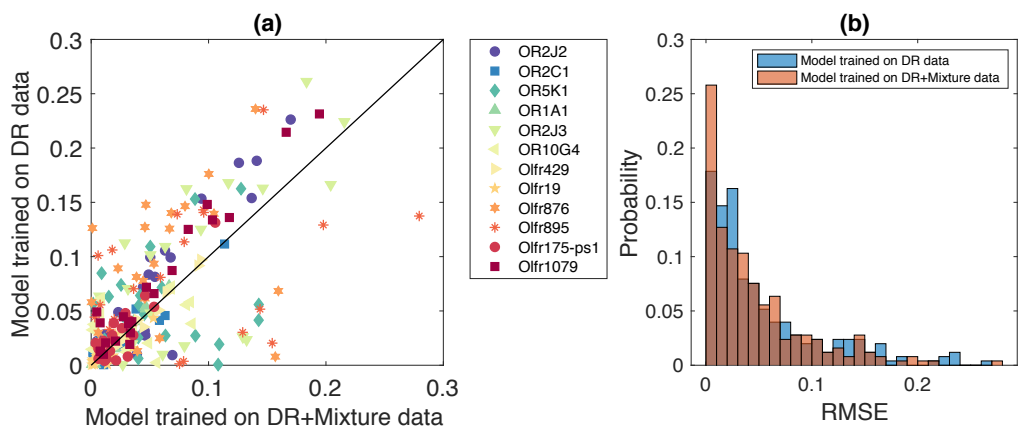

**Fig. S6. Comparison of RMSE for model trained on dose-response and mixture supplemented dose-response data:** (a) RMSE for a model trained only on dose response (DR) data v/s the model trained on both dose-response and mixture data. For mixture constrained training, the responses to three binary mixtures were included in the training set while three single odorant data points were removed. The test set had both binary and ternary mixtures. The mixture-constrained model has lower error. (b) Histogram of RMSE for a model trained only on dose response data and the model trained on both dose-response and mixture data. Training on limited data produces a modest improvement in model predictions.

**Table S1. Binary-Ternary Mixture Compositions (concentrations in $\mu$M)**

| Mixture Number | eugenol | acetophenone | coumarin |
|---|---|---|---|
| 1 | 100 | 100 | 0 |
| 2 | 100 | 0 | 100 |
| 3 | 0 | 100 | 100 |
| 4 | 50 | 50 | 0 |
| 5 | 50 | 0 | 50 |
| 6 | 0 | 50 | 50 |
| 7 | 100 | 50 | 0 |
| 8 | 100 | 0 | 50 |
| 9 | 50 | 100 | 0 |
| 10 | 0 | 100 | 50 |
| 11 | 0 | 50 | 100 |
| 12 | 50 | 0 | 100 |
| 13 | 150 | 150 | 150 |
| 14 | 100 | 100 | 100 |
| 15 | 50 | 50 | 50 |
| 16 | 50 | 100 | 150 |
| 17 | 50 | 150 | 100 |
| 18 | 100 | 50 | 150 |
| 18 | 100 | 50 | 150 |
| 19 | 100 | 150 | 50 |
| 20 | 150 | 100 | 50 |
| 21 | 150 | 50 | 100 |
| 22 | 0 | 0 | 0 |

**Table S2. List of Odorants Used In The Experiments**

| Odorant | CAS | SMILE | Catalog Number |
|---|---|---|---|
| eugenol | 97-53-0 | COC1=C(C=CC(=C1)CC=C)O | E51791-100G |
| acetophenone | 98-86-2 | CC(=O)C1=CC=CC=C1 | A10701-100ML |
| coumarin | 91-64-5 | C1=CC=C2C(=C1)C=CC(=O)O2 | C4261-50G |
| 1-heptanol | 111-70-6 | CCCCCCCO | H2805-250ML |
| 2,3-hexanedione | 3848-24-6 | CCCC(=O)C(=O)C | W255801-SAMPLE-K |
| 2-heptanone | 110-43-0 | CCCCCC(=O)C | W254401-SAMPLE-K |
| 2-hexanone | 591-78-6 | CCCCC(=O)C | 103004-10G |
| 3-heptanone | 106-35-4 | CCCCC(=O)CC | W254509-SAMPLE-K |
| 4-chromanone | 491-37-2 | C1COC2=CC=CC=C2C1=O | 122351-10G |
| (-)-carvone | 6485-40-1 | CC1=CCC(CC1=O)C(=C)C | 124931-5ML |
| (+)-carvone | 2244-16-8 | CC1=CCC(CC1=O)C(=C)C | 22070-25ML |
| (+)-dihydrocarvone | 5524-05-0 | CC1CCC(CC1=O)C(=C)C | 37275-25ML |
| dihydrojasmone | 1128-08-1 | CCCCCC1=C(CCC1=O)C | W376302-SAMPLE-K |
| benzyl acetate | 140-11-4 | CC(=O)OCC1=CC=CC=C1 | B15805-100G |
| prenyl acetate | 1191-16-8 | CC(=CCOC(=O)C)C | W420201-SAMPLE-K |
| nonanedioic acid | 123-99-9 | C(CCCC(=O)O)CCCC(=O)O | 246379-100G |

Table S3. 12-Component Mixture Compositions (concentrations in $\mu$M)

| Mixture Number | 1-heptanol | 2,3-hexanedione | 2-hexanone | 2-heptanone | 3-heptanone | 4-chromanone | (-)-carvone | (+)-carvone | (+)-dihydrocarvone | dihydrojasmone | benzyl acetate | prenyl acetate |
|---|---|---|---|---|---|---|---|---|---|---|---|---|
| 1 | 100 | 100 | 100 | 100 | 100 | 100 | 100 | 100 | 100 | 100 | 100 | 100 |
| 2 | 33 | 33 | 33 | 33 | 33 | 33 | 33 | 33 | 33 | 33 | 33 | 33 |
| 3 | 11 | 11 | 11 | 11 | 11 | 11 | 11 | 11 | 11 | 11 | 11 | 11 |
| 4 | 3.7 | 3.7 | 3.7 | 3.7 | 3.7 | 3.7 | 3.7 | 3.7 | 3.7 | 3.7 | 3.7 | 3.7 |
| 5 | 1.2 | 1.2 | 1.2 | 1.2 | 1.2 | 1.2 | 1.2 | 1.2 | 1.2 | 1.2 | 1.2 | 1.2 |
| 6 | 0.4 | 0.4 | 0.4 | 0.4 | 0.4 | 0.4 | 0.4 | 0.4 | 0.4 | 0.4 | 0.4 | 0.4 |
| 7 | 3 | 1 | 3 | 1 | 0.1 | 1 | 3 | 0.1 | 1 | 0.1 | 0.1 | 1 |
| 8 | 3 | 1 | 3 | 1 | 1 | 3 | 0.1 | 0.1 | 3 | 3 | 0.1 | 1 |
| 9 | 0.1 | 3 | 1 | 0.1 | 3 | 3 | 1 | 1 | 0.1 | 0.1 | 1 | 1 |
| 10 | 0.1 | 1 | 0.1 | 3 | 0.1 | 1 | 1 | 1 | 1 | 1 | 3 | 0.1 |
| 11 | 1 | 1 | 3 | 3 | 1 | 3 | 3 | 1 | 10 | 3 | 10 | 1 |
| 12 | 3 | 1 | 1 | 10 | 10 | 1 | 3 | 1 | 10 | 1 | 10 | 1 |
| 13 | 10 | 1 | 1 | 1 | 10 | 3 | 3 | 1 | 1 | 10 | 3 | 1 |
| 14 | 10 | 10 | 10 | 3 | 10 | 30 | 30 | 10 | 3 | 10 | 3 | 3 |
| 15 | 3 | 3 | 10 | 3 | 30 | 3 | 3 | 30 | 30 | 10 | 3 | 3 |
| 16 | 10 | 30 | 10 | 3 | 30 | 10 | 10 | 10 | 10 | 10 | 10 | 3 |
| 17 | 10 | 3 | 10 | 1 | 1 | 3 | 10 | 0.1 | 0.1 | 1 | 3 | 10 |
| 18 | 1 | 3 | 10 | 1 | 0.1 | 1 | 0.1 | 3 | 0.1 | 0.1 | 3 | 3 |
| 19 | 1 | 3 | 10 | 3 | 1 | 3 | 0.1 | 3 | 1 | 0.1 | 1 | 1 |
| 20 | 0.1 | 0.1 | 10 | 3 | 10 | 1 | 10 | 0.1 | 10 | 1 | 0.1 | 10 |
| 21 | 1 | 0.1 | 10 | 1 | 10 | 3 | 1 | 10 | 0.1 | 3 | 30 | 1 |
| 22 | 3 | 30 | 1 | 30 | 0.1 | 30 | 1 | 30 | 30 | 0.1 | 30 | 1 |
| 23 | 10 | 30 | 3 | 30 | 0.1 | 0.1 | 30 | 0.1 | 10 | 30 | 0.1 | 3 |
| 24 | 1 | 10 | 1 | 30 | 0.1 | 30 | 30 | 0.1 | 0.1 | 3 | 0.1 | 30 |
| 25 | 0 | 0 | 0 | 0 | 0 | 0 | 0 | 0 | 0 | 0 | 0 | 0 |

**Table S4. Parameters of The Competitive Binding Model**
**15 receptors used in binary/ternary analysis**

| Receptor Name | eugenol $F_{max}$*e | eugenol $\log_{10}(EC50)$ | acetophenone $F_{max}$*e | acetophenone $\log_{10}(EC50)$ | coumarin $F_{max}$*e | coumarin $\log_{10}(EC50)$ |
|---|---|---|---|---|---|---|
| OR2J2 | 1.148 | -3.588 | 2.870 | -2.542 | 1.562 | -3.582 |
| OR2C1 | 2.887 | -2.151 | 3.980 | -1.898 | 5.553 | -2.311 |
| OR5K1 | 0.330 | -3.960 | 1.324 | -2.932 | 0.764 | -3.454 |
| OR1A1 | 6.554 | -2.112 | 1.140 | -2.090 | 1.374 | -2.453 |
| OR2J3 | 0.947 | -3.864 | 1.370 | -2.869 | 1.305 | -3.684 |
| OR10G4 | 0.523 | -4.957 | 0.116 | -2.876 | 0.252 | -3.120 |
| Olfr429 | 0.827 | -2.812 | 5.074 | -2.247 | 0.585 | -3.903 |
| Olfr19 | 0.655 | -2.524 | 2.638 | -2.206 | 0.273 | -2.926 |
| Olfr876 | 0.671 | -2.352 | 4.563 | -2.628 | 8.978 | -2.175 |
| Olfr895 | 0.464 | -3.391 | 7.039 | -2.345 | 1.512 | -3.756 |
| Olfr175-ps1 | 1.932 | -2.989 | 2.570 | -2.459 | 0.108 | -3.582 |
| Olfr1079 | 1.240 | -3.112 | 3.143 | -2.747 | 2.577 | -2.933 |
| OR5P3 | 0.541 | -2.818 | 9.142 | -2.555 | 1.218 | -5.328 |
| Olfr1062 | 2.936 | -2.227 | 1.637 | -3.461 | 1.273 | -3.828 |
| Olfr1104 | 1.081 | -2.904 | 3.126 | -2.692 | 1.505 | -3.528 |

**Table S5. Parameters of The Competitive Binding Model**
**12-component mixture analysis**

| Odorant | Olfr168 $F_{max}$*e | Olfr168 $\log_{10}(EC50)$ |
|---|---|---|
| 1-heptanol | 1.287 | -3.225 |
| 2,3-hexanedione | 0.688 | -5.021 |
| 2-hexanone | 1.129 | -5.302 |
| 2-heptanone | 0.807 | -5.140 |
| 3-heptanone | 0.745 | -5.433 |
| 4-chromanone | 0.715 | -3.830 |
| (-)-carvone | 1.856 | -3.844 |
| (+)-carvone | 0.899 | -4.164 |
| (+)-dihydrocarvone | 0.847 | -3.766 |
| dihydrojasmone | 0.822 | -4.273 |
| benzyl acetate | 1.963 | -5.364 |
| prenyl acetate | 0.845 | -5.506 |